\documentstyle[prb,twocolumn,epsf,aps]{revtex}
\begin{document}
\draft \preprint{} \twocolumn[\hsize\textwidth\columnwidth\hsize\csname
@twocolumnfalse\endcsname
\title{Microscopic analysis of shot-noise suppression
in nondegenerate diffusive conductors}
\author{T. Gonz\'alez,
J. Mateos, and D. Pardo}
\address{
Departamento de F\'{\i}sica Aplicada, Universidad de Salamanca, Plaza de
la Merced s/n, E-37008 Salamanca, Spain}
\author{O.M. Bulashenko}
\address{
Departament de F\'{\i}sica Fonamental, Universitat de Barcelona, Av.
Diagonal 647, E-08028 Barcelona, Spain} \author{L. Reggiani}
\address{
Istituto Nazionale di Fisica della Materia, Dipartimento di Scienza dei
Materiali, Universit\`a di Lecce \\ Via Arnesano, 73100 Lecce, Italy}
\date{\today}
\maketitle
\begin{abstract}
{We present a theoretical investigation of shot-noise suppression due to
long-range Coulomb interaction in nondegenerate diffusive conductors.
Calculations make use of an ensemble Monte Carlo simulator
self-consistently coupled with a one-dimensional Poisson solver. We
analyze the noise in a lightly doped active region surrounded by two
contacts acting as thermal reservoirs. By taking the doping of the
injecting contacts and the applied voltage as variable parameters, the
influence of elastic and inelastic scattering in the active region is
investigated. The transition from ballistic to diffusive transport regimes
under different contact injecting statistics is analyzed and discussed.
Provided significant space-charge effects take place inside the active
region, long-range Coulomb interaction is found to play an essential role
in suppressing the shot noise at $qU \gg k_BT$. In the elastic diffusive
regime, momentum space dimensionality is found to modify the suppression
factor $\gamma$, which within numerical uncertainty takes values
respectively of about 1/3, 1/2 and 0.7 in the 3D, 2D and 1D cases. In the
inelastic diffusive regime, shot noise is suppressed to the thermal
value.}
\end{abstract}
 \vskip2pc]
\narrowtext
\section{Introduction}
Shot noise is caused by the randomness in the flux of carriers crossing
the active region of a given device, and is associated with the
discreteness of the electric charge.\cite{ziel70,ziel86,kogan96} At low
frequency (small compared to the inverse of the transit time through the
active region, $f\ll 1/\tau_T$, but sufficiently high to avoid $1/f$
contributions) the power spectral density of shot noise is given by
$S_I(0)=\gamma 2qI$, where $I$ is the dc current, $q$ the electron charge,
and $\gamma$ the suppression factor. Uncorrelated carriers exhibiting
Poissonian statistics are known to be characterized by a full shot-noise
power ($\gamma=1$). However, correlations between carriers can reduce the
noise, leading to suppressed shot noise with $\gamma < 1$. Several
interactions and mechanisms can introduce correlations among carriers,
thus giving rise to different levels of suppression, \cite{dejong97} which
can provide valuable information concerning the carrier kinetics inside
the devices not available from dc characteristics or low-frequency
conductance. \cite{landauer98}
\par
In ballistic systems, like vacuum tubes, shot noise is known since the
seminal work of Schottky, \cite{schottky18} and well understood in terms
of the Poissonian statistics of injected carriers. Within this model, shot
noise has been investigated also in several nonuniform devices like
Schottky diodes, p-n junctions, tunnel diodes, etc. \cite{ziel86} In
contrast with these ballistic or quasi-ballistic structures, in
macroscopic devices, where scattering mechanisms with phonons, impurities
and other carriers determine the transport properties, shot noise is not
usually detected and noise levels close to the thermal value are typically
measured (in the frequency range beyond $1/f$ and generation-recombination
contributions).
\par
With the advent in recent years of mesoscopic conductors, shot noise is
receiving renewed attention. In particular, the phenomenon of suppression,
being a signature of correlations among particles, has emerged as a
subject of relevant interest. The suppression has been predicted
theoretically as a consequence of Pauli exclusion principle under strongly
degenerate conditions in very different situations. In ballistic regime,
shot noise is completely suppressed \cite{kulik84,lesovik89} due to the
non-fluctuating occupation number of the incoming states. In a point
contact, a peak in the noise is predicted in between the conductance
plateaus.\cite{lesovik89} In symmetric double-barrier junctions a 1/2
suppression factor has been theoretically explained by different
authors.\cite{chen91,davies92,dejong95,iannacone97} In the case of elastic
diffusive conductors, a 1/3 reduction of the noise has been calculated for
noninteracting electrons,\cite{dejong95,beenakker92,nagaev92} while in the
case of strong electron-electron scattering the value of the suppression
factor is $\sqrt{3}/4$.\cite{nagaev95,kozub95} Finally, when the devices
become macroscopic and inelastic processes are present, like scattering
with phonons, the noise is expected to reduce to the thermal
value.\cite{shimizu92,liu94} Remarkably, many of these predictions have
been experimentally confirmed,
\cite{li90a,li90b,liefrink94,liu95,birk95,reznikov95,kumar96,steinbach96,henny97,schoel97,pothier97,henny98}
thus opening new and interesting perspectives. Within this scenario, the
understanding of the physical mechanisms originating shot noise and its
suppression in mesoscopic conductors, and more generally in
small-dimensional devices, is a field of growing interest.
\par
Most of the theoretical works carried out so far have considered
degenerate conductors, where the Pauli exclusion principle plays a major
role, and neglected long-range Coulomb interaction among carriers. The
influence of this interaction is known to be relevant to the noise
reduction since the times of vacuum tubes, \cite{north40} and its possible
role in the case of mesoscopic samples \cite{buttiker95} has been
repeatedly claimed by Landauer.
\cite{landauer91,martin92,landauer93,landauer96} However, only recently
some works, always dealing with degenerate conductors, include explicitly
long-range Coulomb interaction\cite{green98} and analyze its influence on
the high-frequency spectrum of shot noise.
\cite{naveh97,nagaev98a,naveh98a,naveh98b,nagaev98b} A systematic analysis
of shot-noise suppression in {\it nondegenerate} conductors with the
inclusion of Coulomb correlations is still lacking.
\par
Coulomb interaction can affect noise in two main ways. On the one hand,
being responsible for total current conservation, it leads to local
voltage fluctuations which try to preserve charge neutrality and as such
may influence the current noise. \cite{beenakker92,buttiker95} On the
other hand, being a repulsive force, it tends to space electrons more
regularly than a Poissonian statistics, thus reducing the possible noise
present in the current flux.\cite{landauer96} This regulation of the
electron motion is particularly evident in the case of ballistic
transport, which was investigated under nondegenerate conditions in
previous works.\cite{shot1,shot3,shot5} However, to our opinion, the role
of Coulomb interaction in suppressing shot noise in the presence of
scattering is still not well assessed. Here two issues are of main
concern: (i) the determination of the reduction value under elastic
diffusive conditions, and (ii) the understanding of the progressive
disappearance of shot noise when passing from the mesoscopic to the
macroscopic realm of conduction under the influence of inelastic
processes.
\par
The aim of this paper is to shed new light on the above issues by
investigating microscopically shot noise and its suppression in {\it
nondegenerate conductors in the presence of elastic and inelastic
scattering and long-range Coulomb interaction}. Our approach differs from
those typically used to analyze noise in mesoscopic systems. Calculations
are based on an ensemble Monte Carlo (MC) simulation self-consistently
coupled with a Poisson solver (PS). Here the scattering mechanisms and the
fluctuations of the self-consistent potential are intrinsically accounted
for. In addition, the approach can analyze different voltage-bias
conditions ranging from thermal equilibrium to high electric fields
necessary for shot noise to manifest without the difficulties that other
methods meet.\cite{green98} With this approach we investigate shot-noise
suppression under the following conditions: (i) crossover between
ballistic and diffusive transport regimes and different carrier injecting
statistics, (ii) diffusive transport regime under elastic or inelastic
scattering, and (iii) 3D, 2D and 1D momentum space. A MC simulation of
shot-noise suppression in a 1D mesoscopic conductor has been performed in
Ref.~\onlinecite{liu97}. It considered a degenerate sample in the presence
of elastic and inelastic scattering where Pauli exclusion principle was
taken into account but Coulomb interaction was neglected.
\par
Despite the fact that the physical system we analyze is nondegenerate,
some of the results we get coincide with those obtained in degenerate
systems by using other approaches. In particular, it is specially
surprising that we find the same {\it universal} 1/3 value for the
shot-noise suppression factor under elastic diffusive regime.\cite{shot4}
This 1/3 value has been obtained by very different theoretical approaches,
going from quantum-phase-coherent \cite{beenakker92} to
semiclassical-diffusive degenerate models.\cite{dejong95,nagaev92,liu97}
The universality of this factor has been demonstrated both in
quantum\cite{nazarov94,blanter97} and semiclassical
contexts.\cite{suk98a,suk98b} In all these cases degenerate conditions are
assumed, and the noise reduction comes from the regulation of electron
motion by the exclusion principle. However, it is not clear if the
different approaches are equivalent, and the reappearance of the 1/3
factor could be just a numerical coincidence, as critically asserted by
Landauer.\cite{landauer96} Here we show another unrelated context where
the 1/3 suppression factor appears, in which neither phase coherence nor
Fermi statistics are present. In our case the origin of the effect is
completely classical, and the correlation between electrons comes just
from their Coulomb repulsion.\cite{shot4} Moreover, we show that the 1/3
value only appears when a 3D momentum space is considered. In the 2D or 1D
cases, different factors are obtained, though the physical mechanism of
suppression remains the same. To this purpose, an analytical theory which
explains this dependence of the suppression factor on the dimension of
momentum space has been recently developed by Beenakker.\cite{beenakker98}
Finally, we will also illustrate the essential role played by Coulomb
interaction for the suppression of shot noise by inelastic scattering, as
already stressed by B\"uttiker.\cite{buttiker95}
\par
The outline of the paper is as follows. In Sec. II we discuss the physical
model used for the structures under analysis, with special emphasis on the
modeling of the contacts. We also provide the details of the MC simulation
and noise calculations. Section III presents the results of shot-noise
suppression with reference to: (i) the crossover between ballistic and
diffusive transport regimes, (ii) the elastic and inelastic diffusive
regime and, (iii) the role of momentum-space dimensionality. In Sec. IV
the main conclusions and future trends are outlined.
\section{Physical model}
In this section we present the details of the structures here analyzed,
the models used in MC simulations and the procedures for noise
calculation.
\subsection{Analyzed structure}
For our analysis we consider the simple structure shown in Fig.~\ref{est}.
It consists of a lightly doped active region of a semiconductor sample
sandwiched between two heavily doped contacts (of the same semiconductor)
which act as thermal reservoirs and inject carriers into the active
region. The sample is assumed to have a transversal size sufficiently
thick to allow a 1D electrostatic treatment in the $x$ direction and
neglect the effects of boundaries in $y$ and $z$ directions. The doping of
the contacts $n_c$ is taken to be much higher than that of the active
region $N_D$. The carrier density at the contacts corresponds to their
doping concentration; all impurities are assumed to be ionized at the
temperature $T=300 \ K$ considered here. The contacts are assumed to have
no voltage drop inside and to remain always at thermal equilibrium.
Accordingly, when a voltage $U$ is applied to the structure, all the
potential drop takes place inside the active region, between the positions
$x=0$ and $x=L$. We shall analyze shot-noise suppression in different
structures of same length but variable contact doping.
\subsection{Contact models}
The modeling of carrier injection from contacts can be crucial for the
noise behavior in mesoscopic devices, especially in the case of ballistic
transport.\cite{shot5} To provide a complete model for the contacts and
thus define the related sources of randomness in the carrier flux, we have
to specify the velocity distribution of the injected carriers
$f_{inj}({\bf v})$, the injection rate $\Gamma$ and its statistical
properties. Here we have denoted ${\bf v} \equiv (v_x,v_y,v_z)$.
\par
Let us consider the process of electron injection from contact 1 into the
active region at $x$=$0$ (see Fig.~\ref{est}). According to the
equilibrium conditions of the contacts, the injected carriers follow a
Maxwellian distribution weighted by the velocity component $v_x$ normal to
the surface of the contact
\begin{equation} \label{distin}
f_{inj}({\bf{v}})=v_x f_{MB}({\bf{v}}), \ \ \ v_x>0,
\end{equation}
where $f_{MB}({\bf{v}})$ is the Maxwell-Boltzmann distribution at the
lattice temperature. The injection rate $\Gamma$, i.e. the number of
carriers per unit time which enter the sample, is given by
\begin{equation} \label{g}
\Gamma=n_c \bar v_+S,
\end{equation}
where $S$ is the cross-sectional area of the device, and
\begin{equation} \label{vel}
\bar v_+=\int_0^{\infty}\int_{-\infty}^{\infty}\int_{-\infty}^{\infty}
f_{inj}({\bf{v}})d v_x d v_y d v_z= \sqrt{\frac {k_BT}{2 \pi m}},
\end{equation}
with $k_B$ the Boltzmann constant and $m$ the carrier effective mass. The
injection rate is taken to be independent of the applied voltage. Due to
the very high value of $n_c$ as compared to $N_D$, any possible influence
of the applied voltage (especially for high values) on the contacts and,
consequently, on the injection rate, is neglected. Thus, the boundary
injecting condition at the contacts is described through the constant
injection rate $\Gamma$. The maximum current that a contact can provide in
the ballistic limit is the saturation current $I_S=q\Gamma$.
\par
According to the nondegenerate distribution of carriers, the random
injection at the contacts is taken to follow a Poissonian statistics.
Thus, the time between the injection of two consecutive electrons,
$t_{inj}$, is generated with a probability per unit time given by
\begin{equation} \label{inj}
P(t_{inj})=\Gamma e^{-\Gamma t_{inj}}.
\end{equation}
In the simulation we make use of Eq.~(\ref{inj}) to generate $t_{inj}$,
which, following the MC technique, is given by
$t_{inj}=-\frac{1}{\Gamma}\ln(r)$, where $r$ is a random number uniformly
distributed between 0 and 1. Electrons are injected at $x=0$ and $x=L$
into the active region of the structure according to the above stochastic
rate. When a carrier exits through any of the contacts it is canceled from
the simulation statistics, which accounts only for carriers that are
inside the active region at the given time $t$. Thus, the number of
carriers in the sample $N(t)$ is a stochastic quantity which fluctuates in
time due to the random injection from the contacts and we can evaluate
both the time-averaged value $\langle N\rangle$ and its fluctuations
$\delta N(t)=N(t)-\langle N\rangle$.
\par
Consistently with the condition assumed here that the doping of the
contacts is much higher than that of the active region, all the built-in
effects associated with the diffusion of carriers around the contacts take
place exclusively in the active region. Therefore, the effects related to
possible charge fluctuations at the contacts are neglected in the
calculation of the current. In any case, these effects are expected to
appear at high frequencies (comparable with those of the plasma) while we
are mostly interested in the low-frequency region of the noise spectrum.
\par
Unless otherwise indicated, calculations shall make use of the above
contact model, which appears to be physically plausible under
nondegenerate conditions. However, to analyze the influence of the contact
injecting statistics on the noise behavior, alternative models are also
used. In particular, for the injected carriers we shall consider: (i)
fixed velocity instead of Maxwellian distribution and, (ii)
uniform-in-time instead of Poissonian injection. In case (i) we consider
the same injection rate $\Gamma$ as in the basic model, but all carriers
are injected with identical $x$-velocity $v_x=\sqrt{\pi k_BT/2m}$, which
corresponds to the average velocity of the injected electrons when they
follow a Maxwellian distribution. In case (ii) carriers are injected into
the active region equally spaced in time at intervals of $1/\Gamma$.
\subsection{Monte Carlo simulation}
The transport analysis is carried out by simulating the carrier dynamics
only in the active region of the structure. The influence of the contacts
is included by means of the stochastic injection rate taking place at
positions $x=0$ and $x=L$ as described previously. Under the action of a
dc applied voltage $U$, the carrier dynamics is simulated by an ensemble
MC technique self-consistently coupled with a PS.\cite{jap93} The
simulation is 1D in real space, the Poisson equation being solved in the
direction $x$ of the applied voltage. Typically, a 3D momentum space is
considered. However, to analyze the influence of dimensionality on the
noise suppression, a 2D and 1D momentum space is also considered in some
specific cases. Since we are interested in analyzing only the effects
related to Coulomb interaction, the electron gas inside the sample is
assumed to be nondegenerate thus excluding any additional correlations due
to Fermi statistics. Carriers move inside the active region according to
the semiclassical equations of motion with a constant effective mass, and
undergo isotropic scattering in momentum space. Under the condition of a
constant applied voltage, the instantaneous current in a one-dimensional
structure is given by \cite{lino92}
\begin{equation} \label{current}
I(t)=\frac{q}{L} \sum_{i=1}^{N(t)}v_{xi}(t),
\end{equation}
where $v_{xi}(t)$ is the velocity component along the field direction of
the $i$-th particle. It must be stressed that, although not explicitly
appearing in Eq.~(\ref{current}), the displacement current, which is of
crucial importance for noise calculations, is implicitly taken into
account by constant voltage conditions.\cite{lino92}
\par
To better analyze the importance of Coulomb correlations, which act by
means of the fluctuations of the self-consistent potential, we provide the
results for two different simulation schemes. The first one uses a {\em
dynamic} PS, which means that the potential is self-consistently updated
by solving the Poisson equation at each time step during the simulation to
account for the fluctuations associated with the long-range Coulomb
interaction. The second scheme uses a {\em static} PS, thus calculating
only the stationary potential profile; i.e., once the steady state is
reached, the PS is switched off, so that carriers move in the {\em frozen}
(non-fluctuating) electric field profile. Both schemes are checked to give
exactly the same steady-state spatial distributions and total current, but
the noise characteristics are different. Of course, the PS scheme which is
physically correct is the dynamic one. The static scheme is just used to
evaluate quantitatively the influence of the potential fluctuations on the
total noise, thus enabling us to separate the contribution belonging to
velocity and number from that due to the self-consistent field.
\par
For the calculations we have used the following set of parameters: $T=300$
K, $m=0.25m_0$, dielectric constant $\varepsilon=11.7\varepsilon_0$
($\varepsilon_0$ being the vacuum permittivity), $L=2000$ \AA, $n_c$
ranging between $10^{13} \ \rm cm^{-3}$ and $10^{18} \ \rm cm^{-3}$ and
$N_D= 10^{11} \ \rm cm^{-3}$. According to Eq.~(\ref{g}), $\Gamma$ is
proportional to $n_c$ and it determines the level of space charge inside
the active region, which will be characterized by the dimensionless
parameter $\lambda$, defined as \cite{shot1,shot5}
\begin{equation} \label{lan}
\lambda=\frac {L}{L_{Dc}},
\end{equation}
where $L_{Dc}=\sqrt{\varepsilon k_BT/q^2n_c}$ is the Debye length
corresponding to the carrier concentration at the contacts. In present
calculations $\lambda$ will take the minimum value of $0.15$ ($n_c=10^{13}
\ \rm cm^{-3}$), for which the effects of Coulomb repulsion between
electrons are practically negligible, and the maximum values of $30.9$
($n_c=4\times10^{17} \ \rm cm^{-3}$) and 48.8 ($n_c=10^{18} \ \rm
cm^{-3}$), for which quite significant electrostatic screening takes
place.
\par
Scattering mechanisms are introduced in the simulation in a simple way by
making use of an energy independent relaxation time $\tau$. Accordingly,
we will consider separately elastic and inelastic interactions, both taken
to be isotropic. In the elastic case, after each scattering event the
direction of the velocity is randomized while the energy is conserved. In
the inelastic case, beside randomizing its velocity, the electron is also
completely thermalized by generating its velocity components after the
scattering in accordance with a Maxwell-Boltzmann distribution at the
lattice temperature. This model of strong inelastic scattering is adopted
to evidence clearly the influence of energy dissipation on the noise
suppression.
\par
While $L$ remains constant, the value of $\tau$ is appropriately varied
from 10 ps to 1 fs to cover both the ballistic and diffusive transport
regimes. The transition between these regimes will be characterized by the
ratio between the carrier mean free path $\ell$, taken as $\ell = 2 \bar
v_+ \tau$, and the sample length $L$. The conditions $\ell /L \gg 1$ and
$\ell /L \ll 1$ correspond to the perfect ballistic and perfect diffusive
regimes, respectively.
\par
Typical values of the time step and number of meshes in real space used
for the PS are 2 fs and 100, respectively, except for the cases when $\tau
< 5$ fs for which the time step is taken of 0.2 fs. As test of numerical
reliability we have checked that by reducing the time step or by
increasing the number of meshes the results remain the same. The average
number of simulated particles in the active region ranges between $50$ and
$2000$ depending on contact doping, transport regime, and applied voltage.
\subsection{Noise calculation}
In the analysis of shot noise we are particularly interested in the
low-frequency value of the spectral density of current fluctuations
$S_I(0)$, for then obtaining the suppression factor $\gamma=S_I(0)/2qI$.
To this end, from the simulation we firstly evaluate the autocorrelation
function of current fluctuations
\begin{equation} \label{ci}
C_I(t)=\langle \delta I(0)\delta I(t)\rangle,
\end{equation}
where $\delta I(t)=I(t)-\langle I\rangle$ is the instantaneous current
fluctuation. Then, the corresponding spectral density is obtained by
Fourier transforming
\begin{equation} \label{si}
S_I(f)=2\int_{-\infty}^\infty C_I(t)e^{i2\pi ft}dt.
\end{equation}
Once steady-state conditions in the active region of the structure are
reached, the simulation typically consists of 350000 time steps. At the
end of each time step the instantaneous current $I(t)$ is calculated. From
these sufficiently long sequence of values of $I(t)$, the time average of
the current $\langle I\rangle$ is determined and the current
autocorrelation function is directly calculated following Eq.~(\ref{ci}).
To clarify the role of number and velocity contributions to the current
noise, the autocorrelation function and the spectral density are
decomposed into three main terms as
\begin{equation} \label{desc1}
C_I(t)=C_V(t)+C_N(t)+C_{VN}(t),
\end{equation}
respectively given by
\begin{mathletters} \label{desc2}
\begin{eqnarray}
C_V(t)&=&\frac{q^2}{L^2} \langle N\rangle^2 \langle \delta v(0) \delta
v(t)\rangle, \\ C_N(t)&=&\frac{q^2}{L^2} \langle v\rangle^2 \langle \delta
N(0) \delta N(t)\rangle, \\ C_{VN}(t)&=&\frac{q^2}{L^2} \langle
v\rangle\langle N\rangle \langle \delta v(0) \delta N(t) \nonumber \\
&&\phantom{xxxxxx} + \delta N(0) \delta v(t)\rangle.
\end{eqnarray}
\end{mathletters}
In the above equations $C_V$ is associated with fluctuations in the mean
carrier velocity $v(t)={1\over N(t)}\sum_{i=1}^{N(t)}v_{xi}(t)$, $C_N$
with fluctuations in the carrier number $N(t)$, and $C_{VN}$ with their
cross correlation.\cite{jap93,lino92} To distinguish between the results
obtained from the static and dynamic PS, we shall denote the corresponding
current spectral densities as $S_I^s$ and $S_I^d$, respectively.
\section{Results}
The main results of the present work are organized as follows. Subsections
A and B pertain to a 3D momentum space while subsection C is devoted to 2D
and 1D momentum spaces. Most of the reported results refer to high values
of the space-charge parameter $\lambda$ (typically $\lambda$=30.9), which
implies significant effects related to long-range Coulomb interaction
present in the active region of the structures.\cite{remark1} The
essential influence of $\lambda$ on the suppression factor $\gamma$ is
analyzed at the end of subsection B. We recall that elastic and inelastic
scattering are considered separately in the simulations. In no case both
types of scattering are taken into account simultaneously.
\subsection{Transition ballistic-diffusive regimes}
The behavior of noise in the crossover from ballistic to diffusive
transport regimes is analyzed under far-from-equilibrium conditions ($U
\gg k_BT/q$), since these are necessary for the manifestation of shot
noise. The dependence of noise on applied voltage for the perfect
ballistic regime was already reported in Ref.~\onlinecite{shot5} and for
the perfect diffusive regime it will be analyzed in next subsection.
\par
Fig.~\ref{var} reports the variance of carrier-number fluctuations
normalized to the average carrier number $\langle \Delta N^2 \rangle/
\langle N \rangle$ as a function of $\ell/L$. The values taken by the
variance are found to be practically independent of the transport regime.
The results of the elastic and inelastic cases are very similar; by
contrast two different values are obtained depending on the scheme of the
PS used in the calculations. When the static PS is considered, it is found
that, within numerical uncertainty,\cite{remark2} $\langle \Delta N^2
\rangle = \langle N \rangle$ in all the range of $\ell/L$ reported.
Therefore, in the absence of dynamic Coulomb correlations, the carrier
number follows a Poissonian statistics. In the perfect ballistic regime
this is a consequence of the injecting statistics at the contacts. In the
diffusive regime this is due to the effect of the randomness introduced by
scattering, independently of its elastic or inelastic property. Because of
the Poissonian statistics observed for the carrier-number fluctuations, no
shot-noise suppression is expected within the static PS scheme. On the
contrary, when the dynamic PS scheme is used, a sub-Poissonian behavior is
clearly observed with $\langle \Delta N^2 \rangle$ being about a factor of
$0.4$ lower than $\langle N \rangle$; thus shot-noise suppression is
expected.
\par
Fig.~\ref{sit} reports $S_I(0)$ normalized to $2qI_S$ under the same
conditions of Fig.~\ref{var}. Here the evolution of the current in terms
of $2qI$ is also shown. This evolution exhibits two limiting behaviors:
saturation at $\ell/L\gtrsim 10^{-1}$, typical of a ballistic or
quasi-ballistic regime; and linear decrease at $\ell/L\lesssim 10^{-2}$ as
$\ell/L \to 0$, typical of a diffusive regime. In both the elastic and
inelastic cases, $S_I(0)$ calculated with the static PS coincides exactly
with $2qI$, \cite{remark3} thus leading to the conclusion that {\em the
stream of uncorrelated carriers, even in the presense of intensive
scattering processes (no matter elastic or inelastic), exhibits full shot
noise behavior like in the pure ballistic transport regime}. On the
contrary, with the dynamic PS $S_I(0)$ is systematically lower than $2qI$,
thus evidencing a suppression effect. Here, as expected, elastic and
inelastic curves recover the same value in the ballistic limit, the
suppression corresponding to that caused by the fluctuations of the
potential barrier near the cathode (induced by the space charge) which
controls the current in this regime.\cite{shot1,shot3,shot5} As the
perfect diffusive regime is approached, the suppression remains active,
more pronounced in the inelastic case, and is related to the joint action
of Coulomb repulsion and the presence of scattering.
\par
We shall now consider the dynamic PS scheme only. Fig.~\ref{gt} reports
$\gamma$ as a function of $\ell /L$ for several values of the applied
voltage. In the perfect ballistic regime the two distinct values of
$\gamma$ found refer to the presence or absence of the potential barrier
related to space charge.\cite{shot5} For $U=40 k_BT/q$ the barrier is
still present and the suppression is important. For $U = 80$ and $100
k_BT/q$ the barrier has already disappeared; accordingly the current
saturates and the suppression factor takes on the full shot-noise value.
When the diffusive regime is achieved, in the elastic case $\gamma$
attains a constant value at further decreasing of $\ell / L$, and takes
the same value of about $1/3$ independently of the applied voltage. On the
contrary, in the inelastic case the higher the applied voltage the lower
the value which $\gamma$ is found to take. Remarkably, the value of $\ell
/ L$ at which $\gamma$ starts decreasing when the ballistic regime is
abandoned is the same in the elastic and inelastic cases for a given
applied voltage ($\ell / L \approx 0.3$ and 0.1, for $80$ and 100
$k_BT/q$, respectively). However, when the diffusive regime is approached,
a lower value of $\ell / L$ is required in the inelastic case with respect
to the elastic case for $\gamma$ to reach a constant value. This behavior
can be explained in terms of the different elastic and inelastic
scattering intensity required by the electron system to achieve a
significant equipartition of energy into the three directions of momentum
space.
\par
To better illustrate the above features, Fig.~\ref{redis} shows the
spatial profiles of the total average energy and its three contributions
in the $x$, $y$ and $z$ directions of momentum space for several values of
$\tau$ in the elastic and inelastic cases at $U=100 k_BT/q$ . For the
longest $\tau$, the total energy practically coincides with the energy in
the $x$ direction independently of elastic and inelastic cases, since the
transport is nearly ballistic. In the case of elastic scattering
[Fig.~\ref{redis}(a)], as $\tau$ is shortened the energy in the $x$
direction decreases while that in the other two directions increases, in
such a way that the total energy remains constant, as follows from elastic
conditions. By further shortening $\tau$, the energy in the three
directions becomes finally the same and equipartition condition is
achieved. In the inelastic case [Fig.~\ref{redis}(b)], equipartition of
energy is also reached, but only at the shortest $\tau$. The fact that
energy equipartition is more easily reached in the elastic than in the
inelastic case is understood as follows. The occurrence of an elastic
scattering provides to the $y$ and $z$ directions a part of the energy
that the carrier initially had in the $x$ direction, so that the energy
tends to equalize in the three momentum components. However, for the case
of an inelastic scattering only the energy in the $x$ direction changes,
since the energy in the $y$ and $z$ directions always correspond to that
of the thermal equilibrium distribution (${1\over2}k_BT$). This means that
for a given applied voltage, a higher number of scattering events (i.e. a
lower value of $\ell / L$) is necessary to achieve energy equipartition in
the inelastic case with respect to the elastic one. Thus, in Fig.
\ref{redis}(a) it can be observed that for $\ell / L=5.38\times10^{-3}$
and an applied voltage of $100 k_BT/q$, in the elastic case an important
isotropic redistribution of energy has already been achieved, so that
$\gamma$ already takes the value of 1/3 [Fig.~\ref{gt}]. However, in the
inelastic case [Fig.~\ref{redis}(b)], for the same value of $\ell/L$ the
energy in the $x$ direction is still significantly higher than that in the
other two directions of momentum space and, as a result, $\gamma$ has not
still reached its constant value (which corresponds to thermal noise, as
will be shown later) and continues decreasing. These results confirm that
thermal conditions, and therefore thermal equilibrium noise, are not
reached until the inelastic scattering time is so short that the energy
gained by electrons in a mean free path is much lower than the thermal
energy, as already indicated by Landauer.\cite{landauer93}
\par
To analyze the role played by the modeling of the contact injection on the
suppression of noise, Fig.~\ref{cont} reports $\gamma$ as a function of
$\ell / L$ calculated using four different models. They combine
Poissonian/uniform injecting statistics and Maxwellian/fixed-velocity
distribution of the injected carriers. The Poissonian-Maxwellian injection
is the basic one used in calculations. In the perfect ballistic regime,
when carrier transport is deterministic, $\gamma$ crucially depends on the
injection model. Thus, in the case of the uniform fixed-velocity model,
when the injection introduces no extra noise in the current flux, $\gamma$
is found to decrease linearly with the increase of $\ell /L$. The noise
does not vanish completely since, unless $\ell/L\rightarrow\infty$, there
is always some probability of undergoing a scattering mechanism. In this
limit, when the noise is produced just by a few scattering events, it is
clearly observed that elastic interactions lead to more important current
fluctuations than inelastic mechanisms. By approaching the perfect
diffusive regime the suppression factor is found to be independent of the
model used. We conclude that {\it the noise in the diffusive regime (and
particularly the $1/$3 suppression value obtained in the elastic diffusive
case) is independent of the carrier injecting statistics}, and it is only
determined by the joint action of scattering mechanisms and Coulomb
correlations.
\subsection{Diffusive regime}
In this section we shall consider scattering times short enough to ensure
a diffusive transport regime ($\ell/L \lesssim 3 \times 10^{-3}$). In this
regime the noise behavior is closely related to the breadth of the
velocity distribution,\cite{nagaev95,landauer93} as shown for the
$x$-velocity component in Fig.~\ref{dist}. In the case of elastic
scattering, the distribution broadens at increasing applied voltages since
there is no energy dissipation. In the inelastic case, the energy is
maximally dissipated by reducing it to the lattice value after each
scattering event; thus a thermal equilibrium Maxwell-Boltzmann
distribution is obtained independently of the applied voltage. In both
cases the distributions are very slightly displaced to positive
velocities, as implied by the presence of a net current flowing through
the structure. We remark that the distributions shown in Fig.~\ref{dist}
refer to all the carriers present inside the sample. In the elastic case,
the local distributions at given positions are found to exhibit a nearly
isotropic but no longer Maxwellian shape, however their spatial
integration over the whole sample length gives the Maxwellian profile
reported in the figure.
\par
Fig.~\ref{sid} shows $S_I(0)$ normalized to $2qI_S$ as a function of the
applied voltage. Here, the $I-U$ characteristic is also shown in terms of
$2qI$. In the wide range of voltages reported in the figure,\cite{remark4}
the $I-U$ curve exhibits a super-linear behavior which is related to the
importance of space-charge effects and the increase of the carrier number
inside the active region with the applied voltage for $qU\gtrsim k_BT$.
For the highest applied voltages it is found $I\propto U^r$, where the
power $r$ is a function of $\lambda$. In particular, $r=1.7$ for
$\lambda=30.9$ (the case shown in the figure) and $r=1.8$ for
$\lambda=48.8$. In what concerns $S_I(0)$, near thermal equilibrium
conditions ($qU<k_BT$), elastic and inelastic cases exhibit the same value
which satisfies Nyquist relation. In the inelastic case, at increasing
voltages $S_I(0)$ remains practically constant, in accordance with the
velocity distributions of Fig.~\ref{dist}(b). The slight increase shown at
the highest voltages is strictly related to the increase of the carrier
number. On the contrary, in the elastic case at high voltages ($qU/k_BT
\gtrsim 10$), when the velocity distribution starts broadening
significantly from its thermal equilibrium shape [Fig.~\ref{dist}(a)],
$S_I(0)$ increases systematically with $U$, its ratio to the current
remaining constant and providing a value of 1/3 for the suppression factor
$\gamma$.
\par
The suppression factor as a function of the applied voltage is shown in
Fig.~\ref{gd}(a) for $\lambda=30.9$, which corresponds to the same
conditions of the previous figure, and in Fig.~\ref{gd}(b) for $\lambda =
48.8$. The behavior of $\gamma$ is quite similar for both values of
$\lambda$. By comparing the results corresponding to elastic and inelastic
scattering we find that at the lowest bias both cases coincide, by
providing the standard thermal noise given by Nyquist relation. On the
contrary, at the highest voltages the elastic case reaches the $1/3$
limiting value, while the inelastic case decreases systematically. In this
latter case the values of $\gamma$ are closely fitted by this simple
expression for $S_I^{inel}(0)$:
\begin{equation} \label{inel}
S_I^{inel}(0)=4K_BTG_0 {\langle N \rangle \over \langle N \rangle_0},
\end{equation}
where
\begin{equation} \label{g0}
G_0={q^2 \langle N \rangle_0 \tau \over m L^2}
\end{equation}
is the conductance and $\langle N \rangle_0$ the average number of
electrons inside the sample, both in the limit of vanishing
bias.\cite{remark5} Thus, in the case of inelastic scattering the spectral
density can be expressed analogously to that of thermal Nyquist noise
modulated by the variation of carrier number at the given voltage, even in
the presence of a high bias and a net current flowing through the
structure. We conclude that inelastic scattering strongly suppresses shot
noise and makes the noise become macroscopic ($\gamma \ll 1$). These
results confirm previous predictions by Shimizu and Ueda,\cite{shimizu92}
Liu and Yamamoto,\cite{liu94} and Nagaev,\cite{nagaev95} with the
important difference that here we consider a nondegenerate classical
conductor. We remark that present findings prove also that the condition
of inelastic scattering alone does not suffice to suppress shot noise; the
presence of the fluctuating self-consistent electric field remaining a
necessary condition. Indeed, as a counter-proof we refer to the
calculations performed with the static PS scheme, where no suppression has
been detected (see Fig.~\ref{sit}). Therefore, as argued by
B\"uttiker,\cite{buttiker95} {\it it is the combination of both Coulomb
interaction and inelastic scattering which leads to the suppression of
shot noise}.
\par
In the elastic case, the values of $\gamma$ are nicely reproduced by the
following expression for $S_I^{el}(0)$:
\begin{equation} \label{el}
S_I^{el}(0)={8\over3}K_BTG_0 {\langle N \rangle \over \langle N
\rangle_0}+{2\over3}qI\coth \left({qU\over2k_BT}\right),
\end{equation}
which is quite similar to that obtained by Nagaev\cite{nagaev92} in a
degenerate context. Eq.~\ref{el} describes the crossover from
thermal-Nyquist noise for $qU\ll k_BT$ (where $S_I^{el}=4k_BTG_0$) to
suppressed shot noise for $qU\gg k_BT$ (where $S_I^{el}={2\over 3} q I $),
thus providing $\gamma=1/3$ for the highest applied voltages. In contrast
with other approaches, \cite{beenakker92,nagaev92,dejong95,liu97} our
results show that, {\it neither phase-coherence\cite{suk98a,suk98b} nor
degenerate statistics are required for the occurrence of suppressed shot
noise in diffusive conductors}, and purely classical physical processes
can lead to the same 1/3 factor.\cite{landauer98}
\par
To illustrate the physical origin of the 1/3 value, Fig.~\ref{gf}(a)
reports a typical spectrum of the suppression factor under elastic
diffusive conditions for static and dynamic PS schemes. Here the current
spectrum is decomposed into velocity, number, and cross-correlation
contributions, $S_I(f)=S_V(f)+S_N(f)+S_{VN}(f)$, according to
Eqs.~(\ref{desc2}). In the static PS scheme the spectrum clearly shows
that all the three terms contribute to $S_I(f)$, and two different time
scales can be identified. The longest one is associated with the transit
time of carriers through the active region $\tau_T \approx 5$ ps, and is
evidenced in the terms $S_N^s(f)$ and $S_{VN}^s(f)$. The shortest one is
related to the relaxation time of elastic scattering $\tau=5$ fs, and is
manifested in $S_V^s(f)$. Remarkably, the velocity contribution yields 1/3
of the full shot-noise value, while the others two terms provide the
remaining 2/3. Thus, in the static PS scheme full shot noise is recovered
as sum of all the three contributions. On the contrary, in the dynamic PS
scheme $S_N^d(f)$ and $S_{VN}^d(f)$ are found to exactly compensate each
other and, as a result, $S_I^d(f)$ coincides with $S_V^d(f)$ in all the
frequency range. Moreover, $S_N^d(f)$ takes values much smaller than
$S_N^s(f)$. The characteristic time scale of $S_N^d(f)$ and $S_{VN}^d(f)$
differs from that found within the static PS scheme, which was related to
the transit time $\tau_T$. Now in the dynamic case it is the dielectric
relaxation time corresponding to the carrier concentration at the contacts
$\tau_d=0.46$ ps which determines the cutoff of the contributions
belonging to number fluctuations. In the frequency range between the
transit and collision frequency values it is interesting to notice that
both static and dynamic PS schemes yield $\gamma = 1/3$, thus relating the
suppression factor to velocity fluctuations only. However, at low
frequencies only the dynamic scheme takes this value by virtue of Coulomb
correlations, which are responsible for the reduction of $S_N^d(f)$ and
the mutual compensation of $S_N^d(f)$ and $S_{VN}^d(f)$ contributions. It
is remarkable that $S_V^s(f)=S_V^d(f)$ in all the frequency range, which
implies that velocity fluctuations are not affected by long-range Coulomb
interaction, but just by scattering mechanisms. Coulomb repulsion affects
only the contributions where carrier-number fluctuations are involved.
Fig.~\ref{gf}(b) reports the spectrum for inelastic scattering. Here, the
same features of the elastic case are observed, with the important
difference that $S_V^d(f)$ is much lower than when there is no energy
dissipation.
\par
So far we have analyzed structures where Coulomb repulsion plays an
important role (i.e. $\lambda \gg 1$). To check to which extent this
interaction is determinant for noise suppression, Fig.~\ref{gl} reports
$\gamma$ and the three contributions in which it has been decomposed as a
function of $\lambda$. Here, we present both the elastic and inelastic
cases calculated under far-from-equilibrium conditions within the dynamic
PS scheme. The variance of the number of carriers inside the sample is
also shown to analyze the evolution of carrier statistics as a function of
$\lambda$. For the lowest values of $\lambda$, when space-charge effects
and in turn Coulomb interaction are negligible, full shot noise is
observed. As $\lambda$ increases, $\gamma$ starts decreasing from unity
until reaching a constant value for $\lambda \gtrsim 30$. It is remarkable
that the contribution of velocity fluctuations to $\gamma$ does not vary
significantly with $\lambda$, being much smaller in the inelastic case
with respect to the elastic one because of the quasi-thermal conditions
imposed by energy dissipation. On the contrary, the contributions
associated with number and velocity-number fluctuations are strongly
affected by the increase of $\lambda$. Indeed, their absolute value
decreases systematically and, being opposite in sign, they compensate each
other at the highest values of $\lambda$, so that $S_I=S_V$. We also
notice that the transition from full shot noise to suppressed shot noise
regimes is sharper in the inelastic case. The variance of the carrier
number inside the sample resembles the behavior of $\gamma$, indicating
that the increase of $\lambda$ leads to more pronounced sub-Poissonian
statistics until reaching the final regime for $\lambda \gg 1$. Therefore,
we conclude that the $1/3$ value exhibits several {\it universal}
properties, namely, it is independent of: (i) the scattering strength
(once $\ell \ll L$); (ii) the applied voltage (once $qU\gg k_BT$), (ii)
the screening length (once $\lambda\gg 1$), and (iv) the carrier injecting
statistics.
\subsection{Dependence on momentum space dimensionality}
The results reported so far refer to a 3D momentum space. In contrast to
degenerate diffusive systems where, provided quasi-one dimensional
conditions in real space are attained, noise suppression is independent of
the number $d$ of momentum space dimensions, an interesting feature of
nondegenerate diffusive systems is that noise suppression can depend on
$d$. For the inelastic case considered here no dependence on $d$ has been
found, since there is no influence of the velocity components transversal
to the electric field direction on transport and noise properties of the
structures. On the contrary, in the elastic case the suppression factor is
found to depend significantly on $d$, \cite{shot4} since the transversal
velocity components constitute a channel for energy redistribution which
affects the transport properties of the structure. Therefore, below we
shall focus our analysis on the elastic case. Accordingly, when $d=2$ in
the simulation the carrier velocity is randomized in two components after
each scattering event, and when $d=1$ the isotropic character of
scattering is accomplished by inverting the carrier velocity with an
average (back-scattering) probability $P_b=0.5$.
\par
Fig.~\ref{dim}(a) reports $\gamma$ as a function of $\ell / L$ for the 1D,
2D and 3D cases at high voltages ($U=40 k_BT/q$) calculated within the
dynamic PS scheme. We notice that, when calculated within the static PS
scheme, the results in 1D and 2D cases do not exhibit any shot-noise
suppression, like in the 3D case. For the highest values of $\ell/L$, in
all three cases $\gamma$ approaches the asymptotic value corresponding to
the ballistic limit ($\gamma=0.045$)\cite{shot1}, where the behavior is
independent of $d$. At a given value of $\ell /L$, a higher deviation from
the asymptotic ballistic value is observed for lower $d$. This is due to
the fact that, in average, elastic interactions introduce higher
fluctuations of the carrier $x$-velocity the lower is the number of
available momentum states after the scattering mechanism (in particular,
just 2 in the 1D case). For this same reason, the increasing presence of
scattering as $\ell /L$ is reduced leads to higher values of the
suppression factor the lower is the dimensionality. Remarkably, within
numerical uncertainty the limit value reached by $\gamma$ in the perfect
diffusive regime is found to be, respectively, of $1/3$, $1/2$, and 0.7
for 3D, 2D, and 1D. \cite{remark6} Fig.~\ref{dim}(b), by reporting
$\gamma$ in the diffusive regime as a function of the applied voltage,
provides evidence that these limit values are independent of the bias once
$q U\gg k_BT$. The origin of the suppression is the same in all three
cases: the joint action of Coulomb correlations and elastic scattering,
which leads to the result $S_I(0)=S_V(0)$ [as shown in Fig.~\ref{gf}(a) in
the 3D case], where $S_V(0)/2qI$ under perfect diffusive regime is a
function of the dimensionality of momentum space.
\par
In a recent work, Beenakker\cite{beenakker98} has developed an analytical
theory able to explain the dependence of $\gamma$ on $d$ in nondegenerate
diffusive conductors and has provided a closed expression for $\gamma$ as
a function of $d$. In particular, the theory predicts for $\gamma$ the
values of $0.34$, $0.51$, and $0.92$ for, respectively, the 3D, 2D and 1D
cases, in agreement (more closely for 3D and 2D) with present findings.
From that analysis, Beenakker concluded that the proximity of $\gamma$ to
$1/d$ is accidental. In view of the simplifications introduced in his
theory we believe that such a conclusion cannot rigorously be applied to
the present model.
\par
For reason of completeness, we have finally analyzed the possible
influence of an anisotropic elastic scattering on transport and noise in
the 1D case. Fig.~\ref{1d} reports $\gamma$ and the current as a function
of $\ell / L$ for three different values of $P_b$: 0.5 (isotropic case),
0.25 and 0.10 (less pronounced back-scattering). The noise results differ
only in the quasi-ballistic regime, where the current remains nearly
constant. Once the diffusive regime is achieved, the current decreases
linearly with $\ell / L$ and $\gamma$ takes the same value of about 0.7
independently of the degree of scattering anisotropy. As expected, we have
found that the smaller the value of $P_b$ the wider the quasi-ballistic
range, and for a given value of $\ell /L$ in this range the stronger the
suppression, as corresponds to closer ballistic conditions.
\section{Conclusions}
We have provided a microscopic analysis of shot-noise suppression in
nondegenerate diffusive conductors. To this purpose, the carrier dynamics
in the active region of a semiconductor structure under the influence of
elastic or inelastic scattering has been simulated by using an ensemble MC
technique self-consistently coupled with a PS. The essential role played
by long-range Coulomb interaction on the shot-noise suppression factor
$\gamma$ has been demonstrated, since no suppression is found in the
absence of the self-consistent potential fluctuations.
\par
We have analyzed shot-noise suppression in the region of crossover from
ballistic to diffusive transport regimes. In the diffusive regime a value
of $\gamma$ independent of sample length is not achieved until a
significant energy redistribution among momentum directions takes place.
For high voltages $qU/k_BT \gg 1$ and long samples $\ell/L \ll 1$, in the
elastic case shot noise is found to be suppressed to a 1/3 value, while in
the inelastic case we have found a stronger suppression the higher the
applied voltage. Noticeably, in the perfect diffusive regime $\gamma$ is
found to be independent of carrier injecting statistics, which implies
that in this regime the noise is just a property of the sample.
\par
Our results show that neither phase coherence nor Fermi statistics are
necessary for the appearance of the 1/3 suppression factor in an elastic
diffusive conductor. In our model, the appearance of this factor requires
the simultaneous fulfillment of the three following conditions: $\ell / L
\ll 1$, $\lambda \gg 1$, and $qU / k_BT \gg 1$. The first implies perfect
diffusive regime, the second strong space-charge effects, and the third
very far-from-equilibrium conditions. Inelastic scattering is found to
further contribute in suppressing shot noise, by reducing it to values
close to thermal Nyquist noise under strong dissipative conditions.
However, for this suppression to take place it is necessary the presence
of long-range Coulomb interaction.
\par
The action of Coulomb repulsion in suppressing shot noise takes place
through the reduction of the contributions associated with carrier-number
fluctuations to the total noise spectral density. In particular, the
compensation between number and velocity-number terms implies that the
total noise is finally determined just by the contribution of velocity
fluctuations.
\par
In the elastic case, $\gamma$ depends on the momentum space
dimensionality, the suppression being less pronounced the lower the
dimension of momentum space.
\par
We believe that the present investigation can provide a stimulus for an
experimental verification of some of the reported results. As a valuable
improvement we aim at including degenerate statistics in a further step.
\section*{Acknowledgments}
We gratefully acknowledge the support from the Direcci\'on General de
Ense\~{n}anza Superior e Investigaci\'on through the project PB97-1331,
the Consejer\'{\i}a de Educaci\'on y Cultura de la Junta de Castilla y
Le\'on through the project SA 11/96, and the Physics of Nanostructures
project of the Italian Ministero dell' Universit\'a e della Ricerca
Scientifica e Tecnologica (MURST).

\begin{figure}
\caption{Schematic drawing of the structure under investigation.
}\label{est} \end{figure}

\begin{figure}
\caption{Variance of carrier number inside the active region normalized to
the average carrier number vs the ballistic parameter $\ell /L$ for an
applied voltage of $U=40 k_BT/q$. Calculations are performed by using
static and dynamic PS schemes, and considering elastic and inelastic
scattering mechanisms.} \label{var}
\end{figure}

\begin{figure}
\caption{Low-frequency spectral density of current fluctuations normalized
to $2qI_S$ vs the ballistic parameter $\ell /L$ for an applied voltage of
$U=40 k_BT/q$. Calculations are performed by using static and dynamic PS
schemes, and considering elastic and inelastic scattering mechanisms. The
dependence of the current through $2qI$ is also plotted for comparison.}
\label{sit}
\end{figure}

\begin{figure}
\caption{Shot-noise suppression factor vs ballistic parameter $\ell /L$
for the cases of elastic and inelastic scattering at different applied
voltages. Calculations are performed by using the dynamic PS
scheme.}\label{gt}
\end{figure}

\begin{figure}
\caption{Spatial profiles along the sample of the total average energy and
its three contributions in the $x, y$ and $z$ directions of momentum space
for several values of the scattering time $\tau$ (and of the associated
$\ell /L$). The applied voltage is $U=100 k_BT/q$. (a) refers to elastic
scattering, (b) to inelastic scattering.}\label{redis}
\end{figure}

\begin{figure}
\caption{Shot-noise suppression factor vs the ballistic parameter $\ell
/L$ for an applied bias of $U=40 k_BT/q$ calculated with different contact
models. Calculations refer to the dynamic PS scheme considering elastic
and inelastic scattering. }\label{cont}
\end{figure}

\begin{figure}
\caption{ Velocity distribution function of carriers inside the active
region of the structure for $\ell /L =1.07\times 10^{-3}$ at several
applied voltages. (a) refers to elastic scattering, (b) to inelastic
scattering. }\label{dist}
\end{figure}

\begin{figure}
\caption{Low-frequency spectral density of current fluctuations normalized
to $2qI_S$ vs applied voltage for $\ell /L=1.07\times 10^{-3}$.
Calculations refer to the dynamic PS scheme considering elastic and
inelastic scattering. The current in terms of $2qI$ is also plotted for
comparison. }\label{sid}\end{figure}

\begin{figure}
\caption{Shot-noise suppression factor vs applied bias calculated with the
dynamic PS scheme considering elastic and inelastic scattering with $\ell
/L=1.07\times 10^{-3}$ at two different levels of space charge: (a)
$\lambda=30.9$ and (b) $\lambda=48.8$. The lines correspond to the
fittings of Eqs.~(\ref{inel}) and (\ref{el}). }\label{gd}
\end{figure}

\begin{figure}
\caption{Spectrum of the shot-noise suppression factor under diffusive
regime ($\ell /L=2.69 \times 10^{-3}$) calculated within static and
dynamic PS schemes for (a) elastic and (b) inelastic cases and an applied
voltage of $U=100 k_BT/q$. Different contributions to the total spectrum
are reported in the figure. }\label{gf} \end{figure}

\begin{figure}
\caption{Shot-noise suppression factor and variance of the carrier number
inside the active region as a function of the characteristic parameter of
space charge $\lambda$ under diffusive regime ($\ell /L=2.69\times
10^{-3}$) calculated within dynamic PS scheme for (a) elastic and (b)
inelastic scattering. Velocity, number, and velocity-number contributions
to the suppression factor are also shown in the figure }\label{gl}
\end{figure}

\begin{figure}
\caption{Shot-noise suppression factor for the cases of 1, 2 and 3
dimensions of momentum space calculated within the dynamic PS scheme for
elastic scattering as a function of: (a) ballistic parameter $\ell /L$
with an applied voltage of $U=40 k_BT/q$ and (b) applied bias $U$ under
diffusive regime ($\ell /L=1.07\times 10^{-3}$). }\label{dim}
\end{figure}

\begin{figure}
\caption{Shot-noise suppression factor and normalized current vs the
ballistic parameter $\ell /L$ calculated within the dynamic PS scheme for
elastic scattering in the case of a 1D momentum space. Different curves
refer to the reported values of the probability of scattering in the
backward direction $P_b$. The applied voltage is $U=40 k_BT/q$.
}\label{1d}
\end{figure}


\begin{thebibliography}{99}
\bibitem{remark1}
In particular, a significant inhomogeneity in the charge distribution is
present, as shown in Ref.~\onlinecite{shot5} in the ballistic case.
\bibitem{remark2}
At decreasing $\tau$ the accuracy of the values of $\langle \Delta N^2
\rangle/\langle N \rangle$ in the static scheme is lower. This explains
the dispersion of the plotted results for $\ell/L< 10^{-2}$. To improve
the accuracy of the results in this range, it would be necessary to
consider much longer (not presently affordable) simulation times.
\bibitem{remark4}
At decreasing voltages the determination of $S_I(0)$ can be carried out
without major difficulties, however the lowest value reported for the
current corresponds to $I/I_S=10^{-4}$. Indeed, below this value the
statistical resolution of the simulation is no longer sufficient and the
values of the current become not reliable. For $qU<k_BT$ we expect that
the $I-U$ characteristic of the structures becomes linear because of the
presence of two injecting contacts.
\bibitem{remark5}
Eq.~(\ref{inel}) describes correctly the thermal noise of the present
structure, since the non-linearity of the $I-U$
characteristic\cite{beenakker98} is taken into account through the factor
$\langle N \rangle /\langle N\rangle _0$, which depends on the applied
voltage.
\bibitem{remark6}
The appearance of the 1/3 factor in the 1D MC simulation of
Ref.~\onlinecite{liu97} does not contradict present results because there
it was due to the effect of Pauli exclusion principle, while here the
correlation between electrons comes from their Coulomb repulsion.
\end{thebibliography}
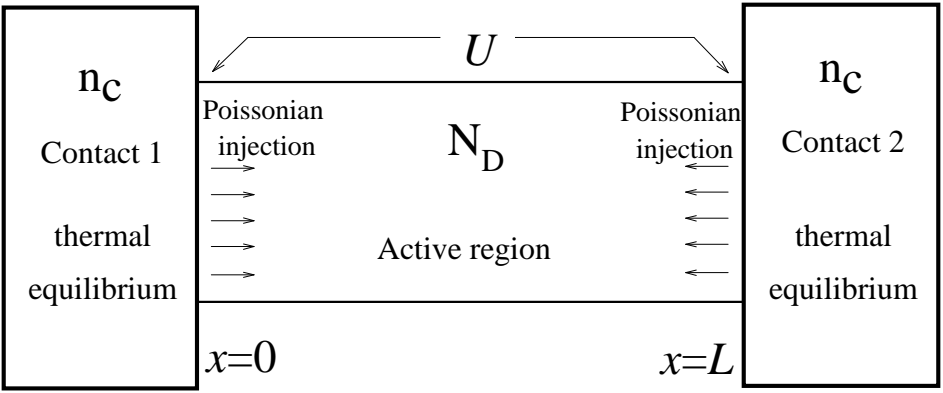
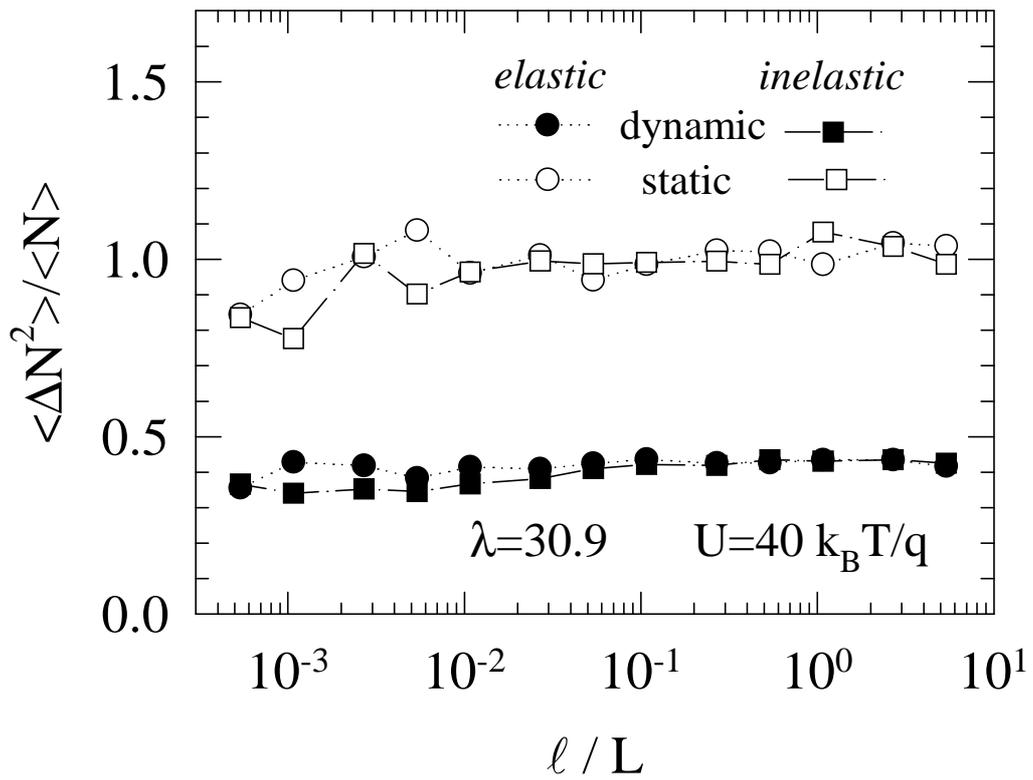
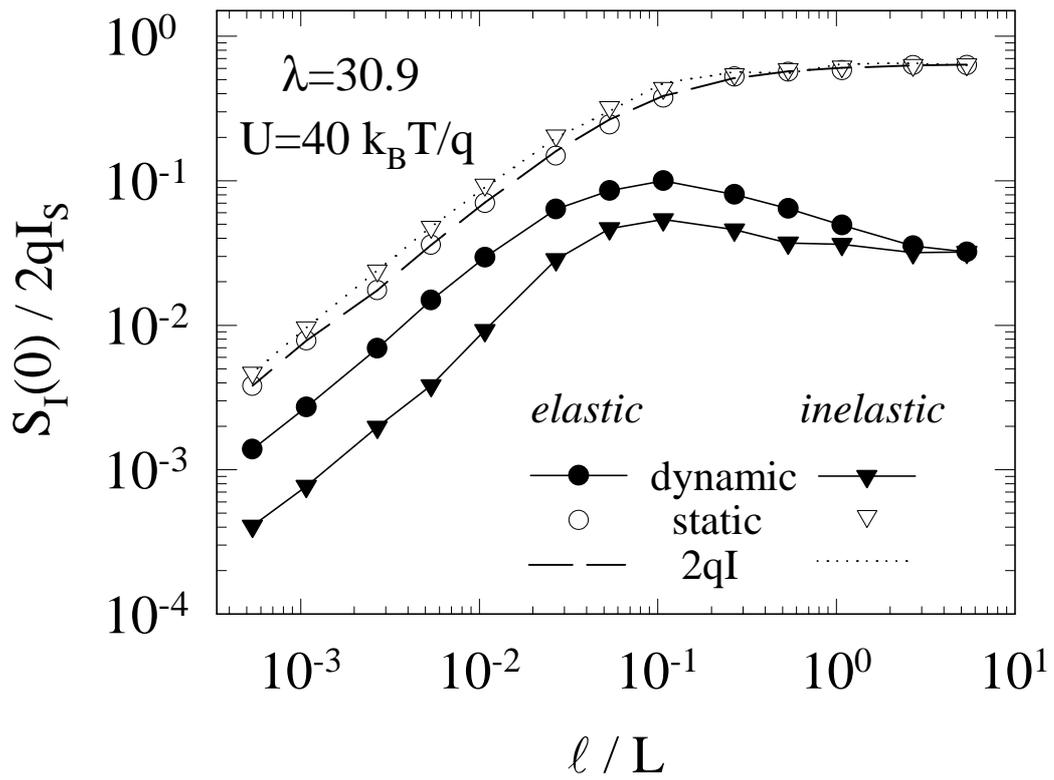
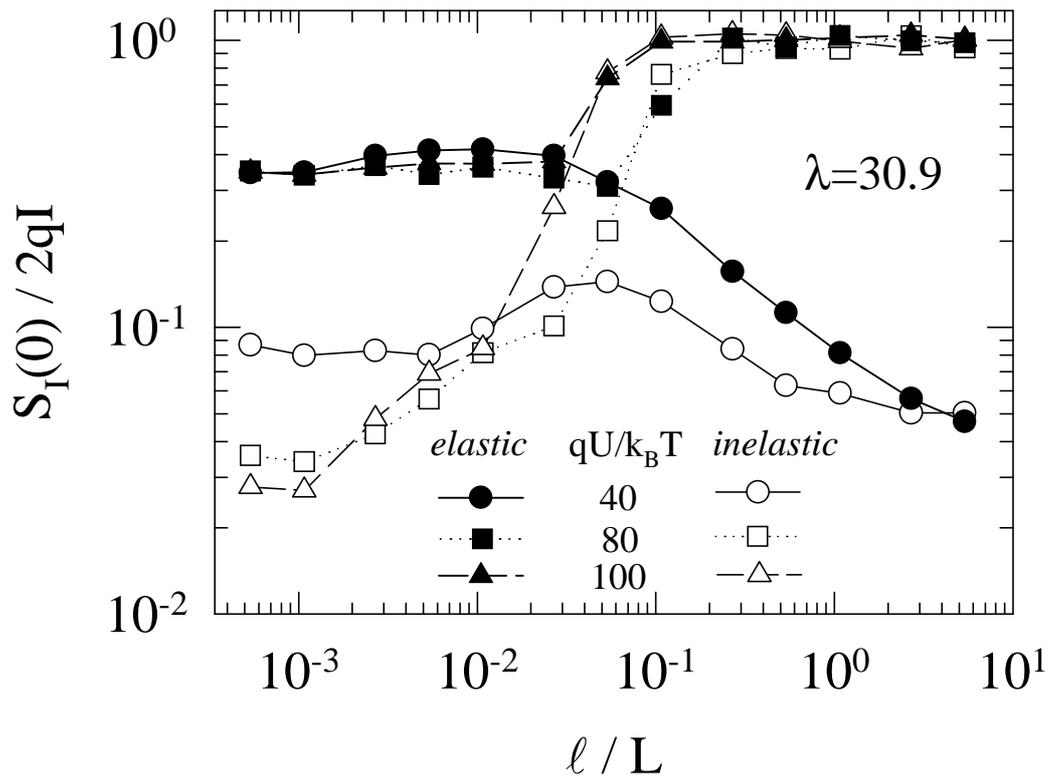
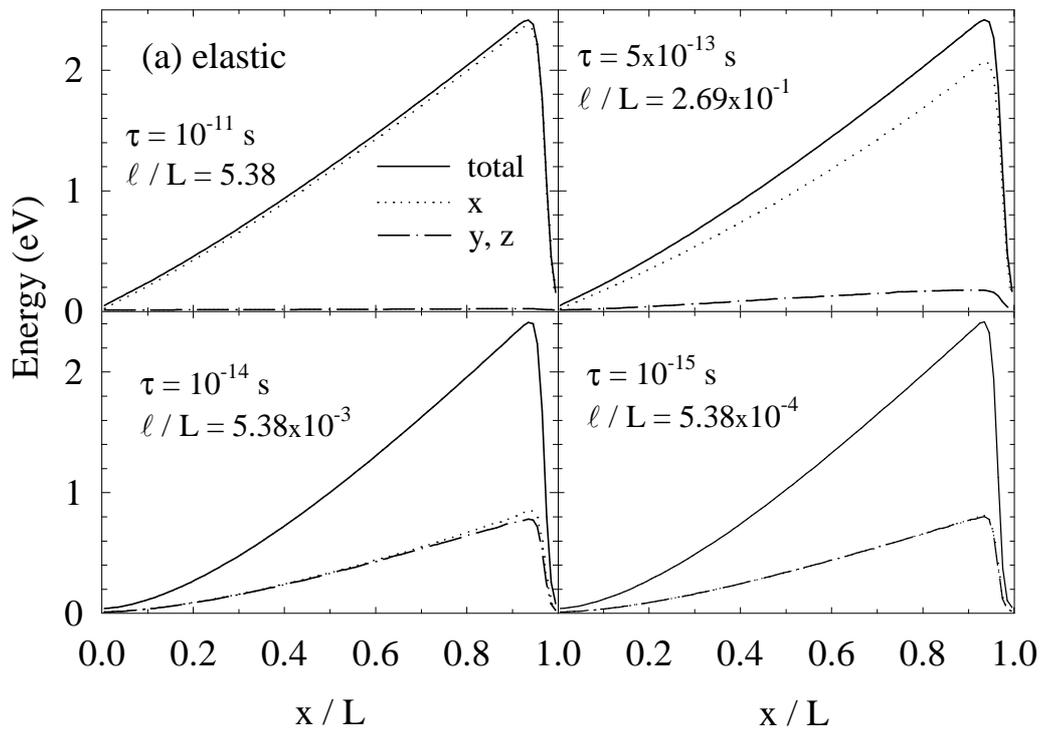
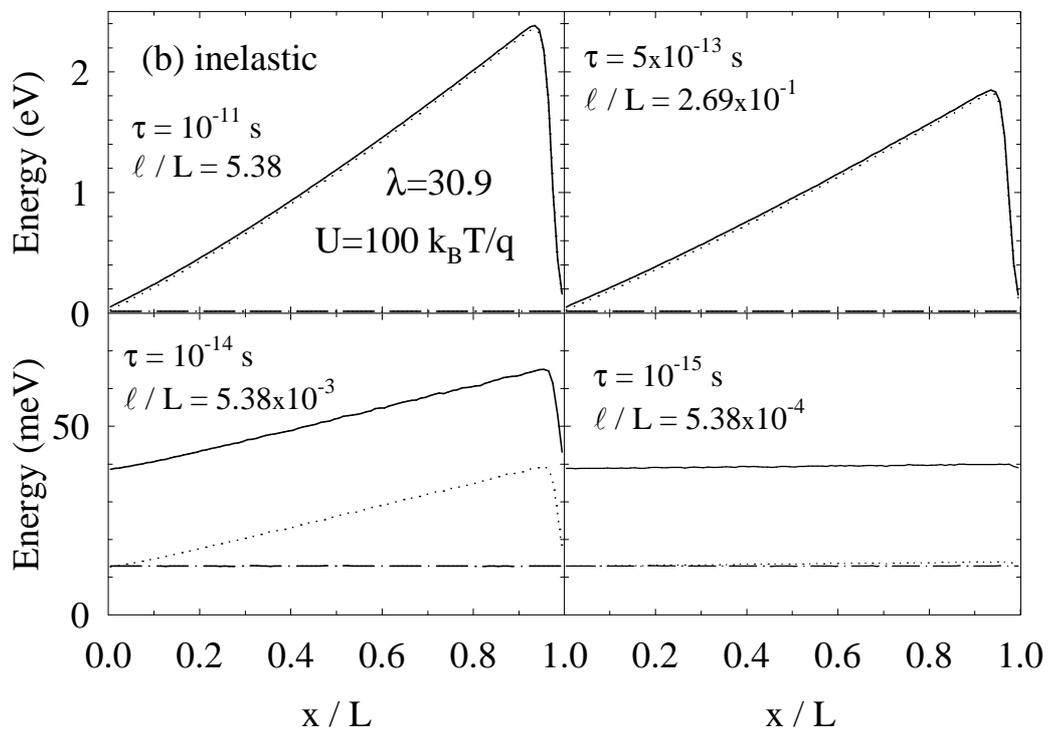
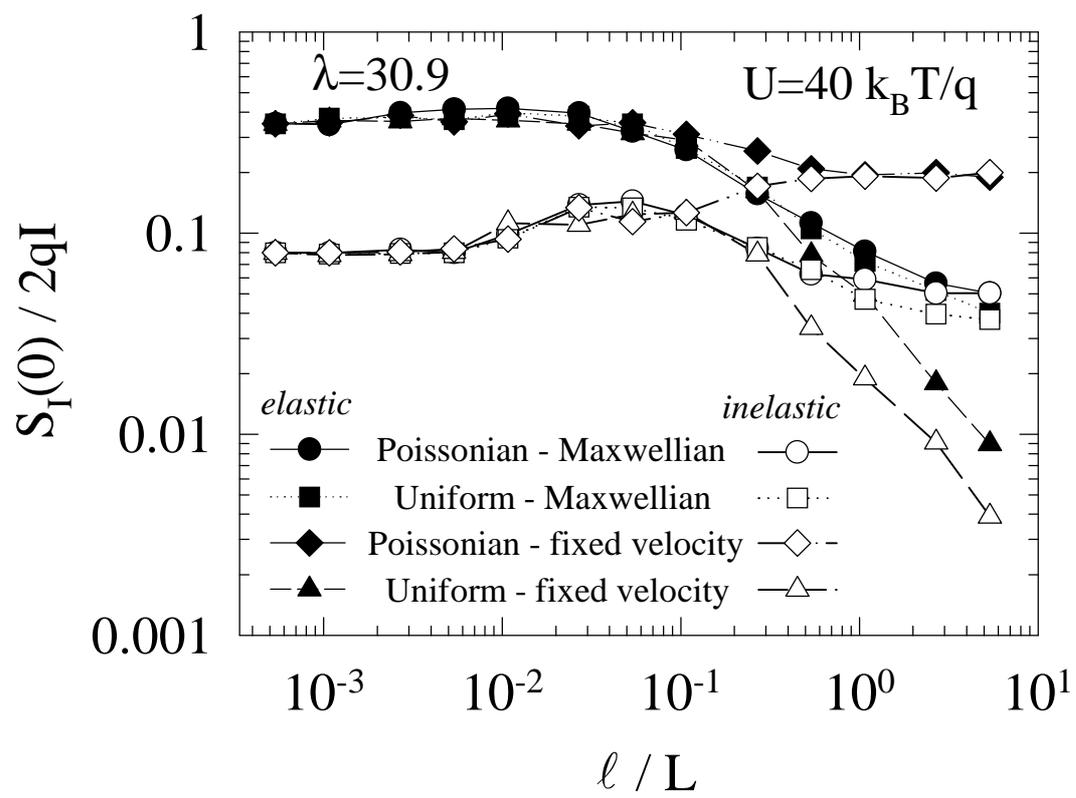
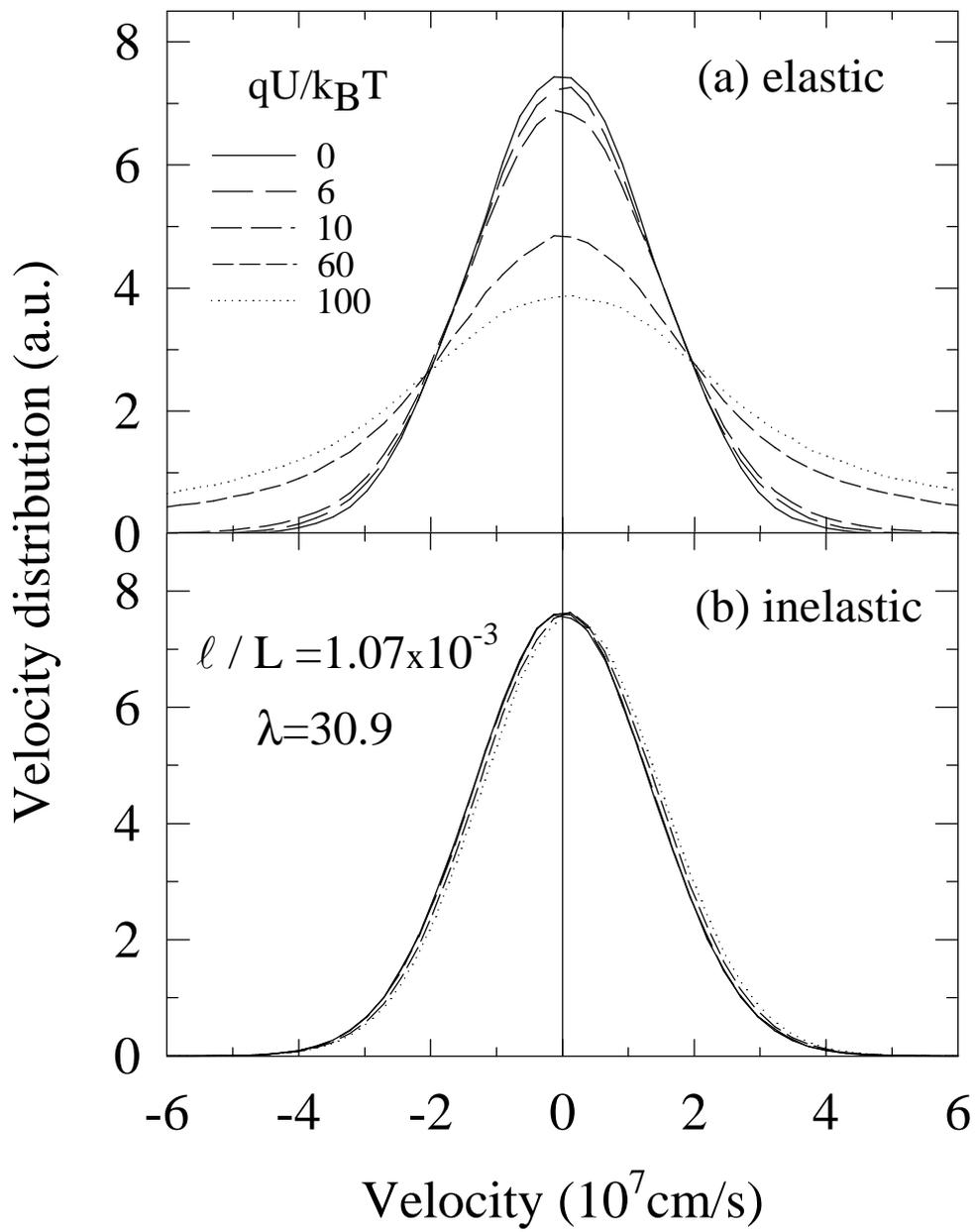
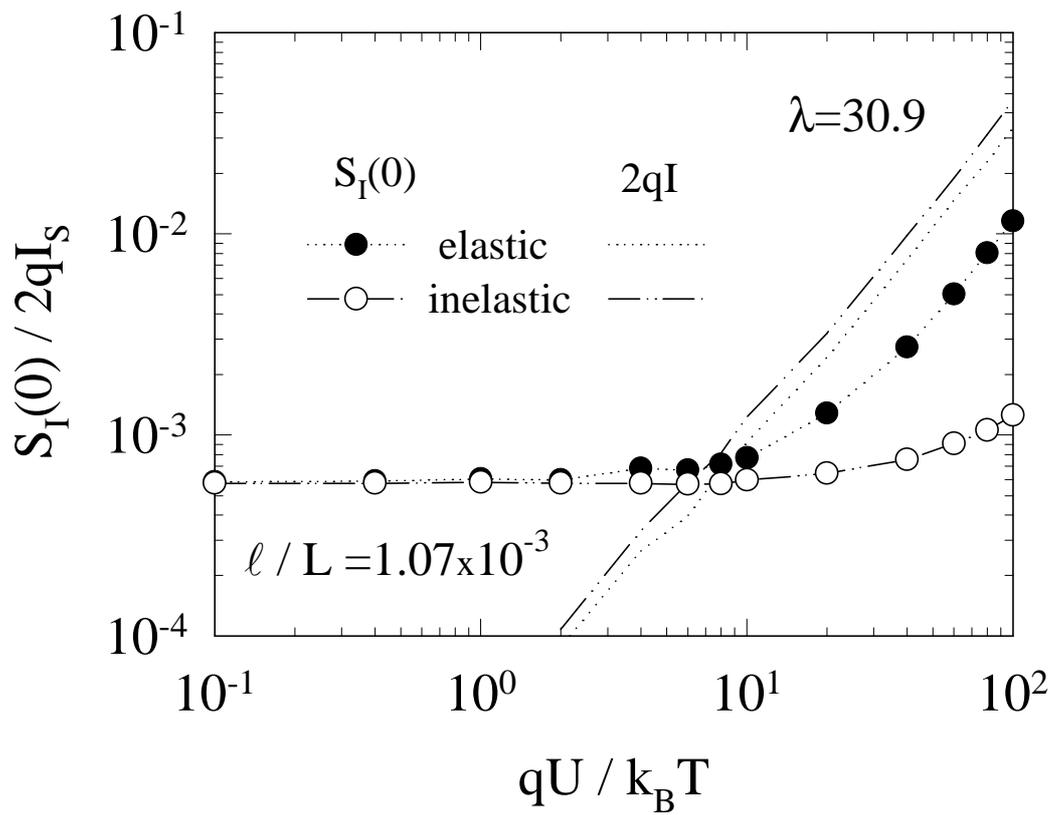
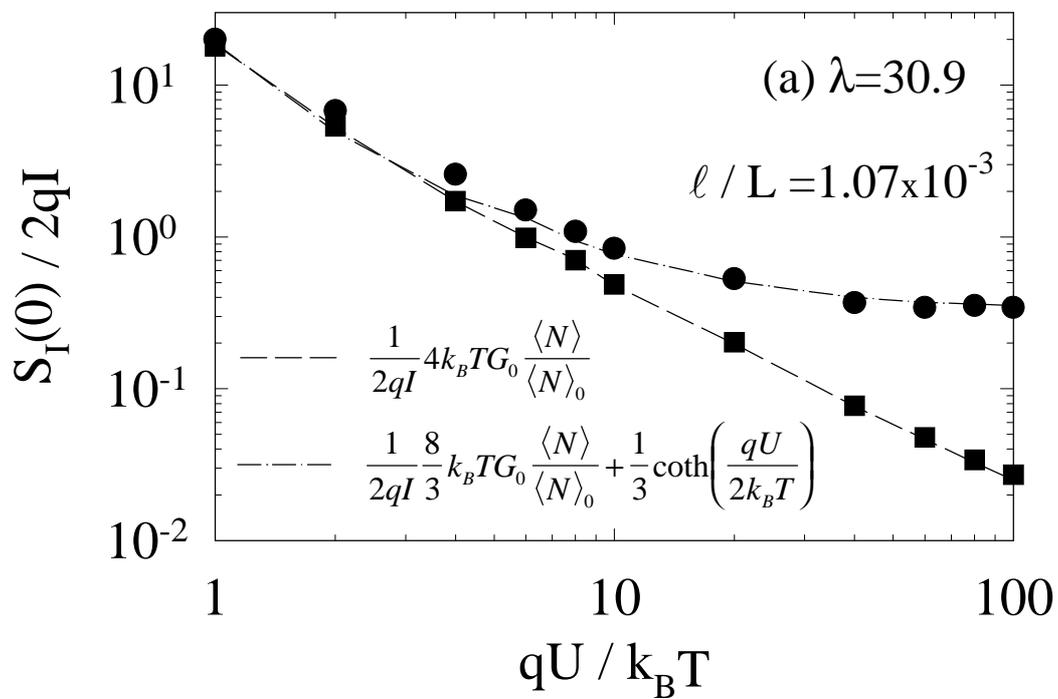
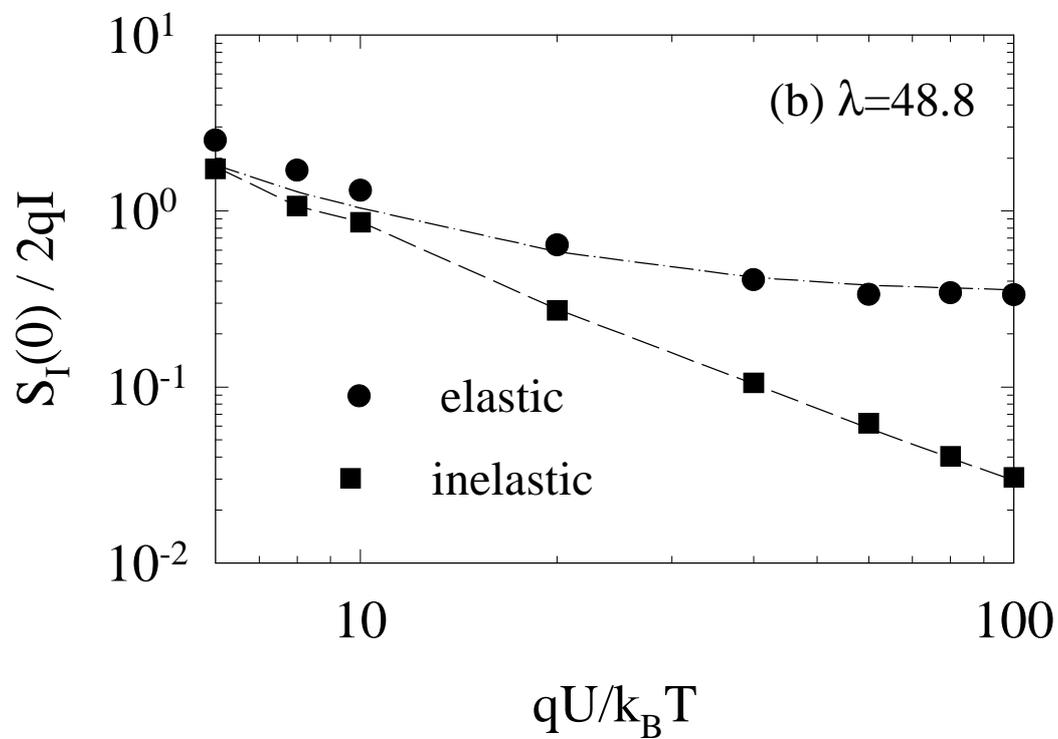
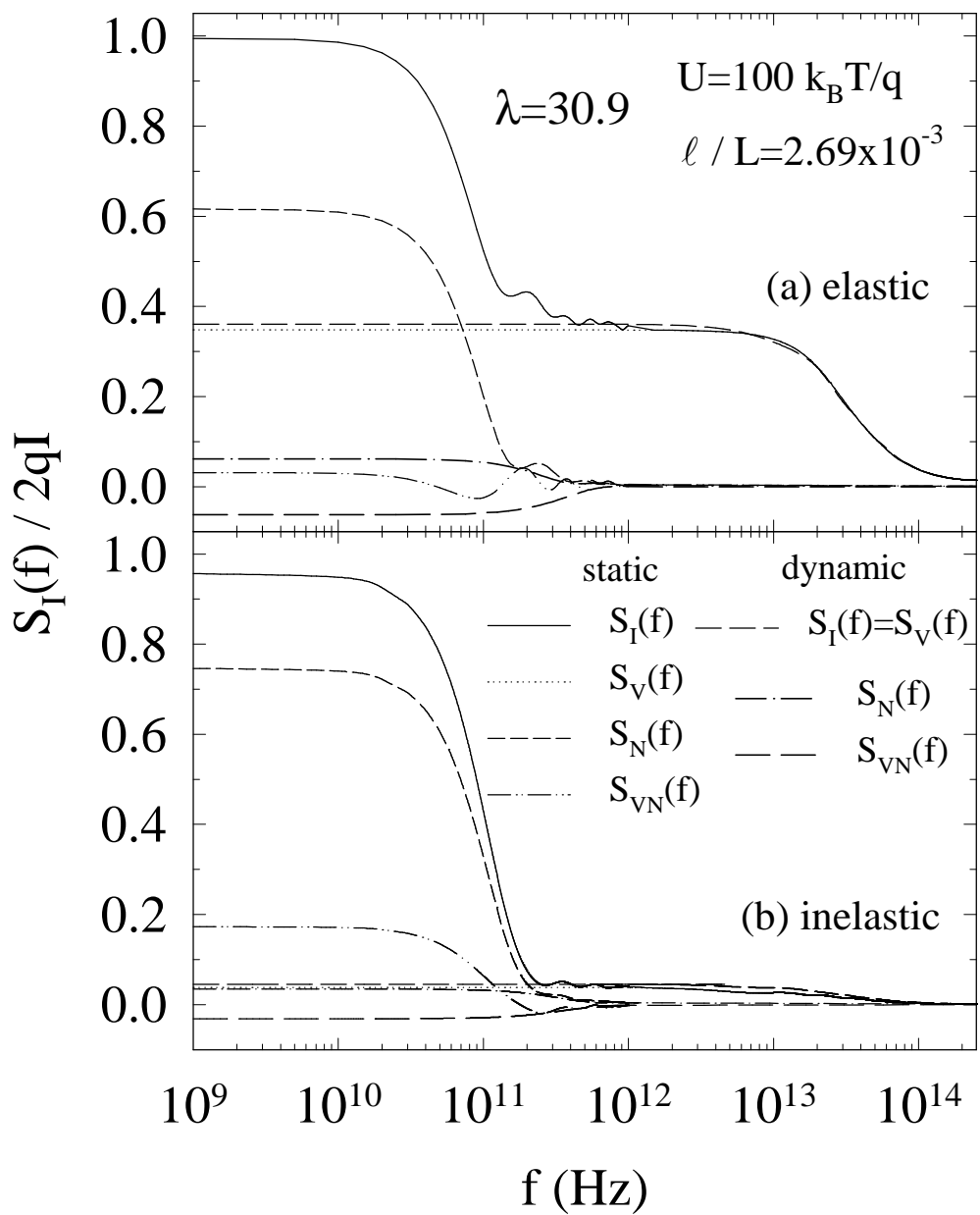
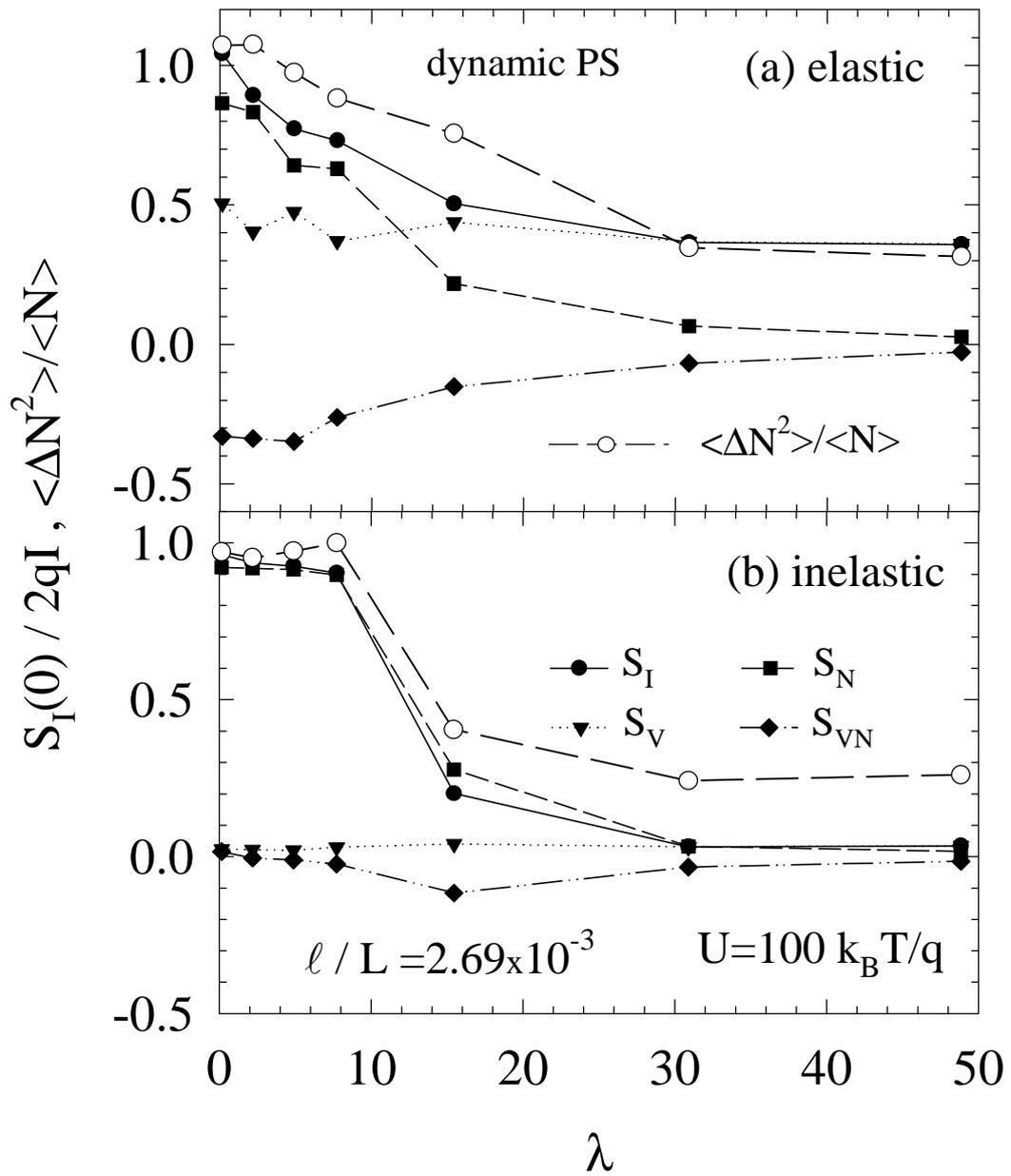
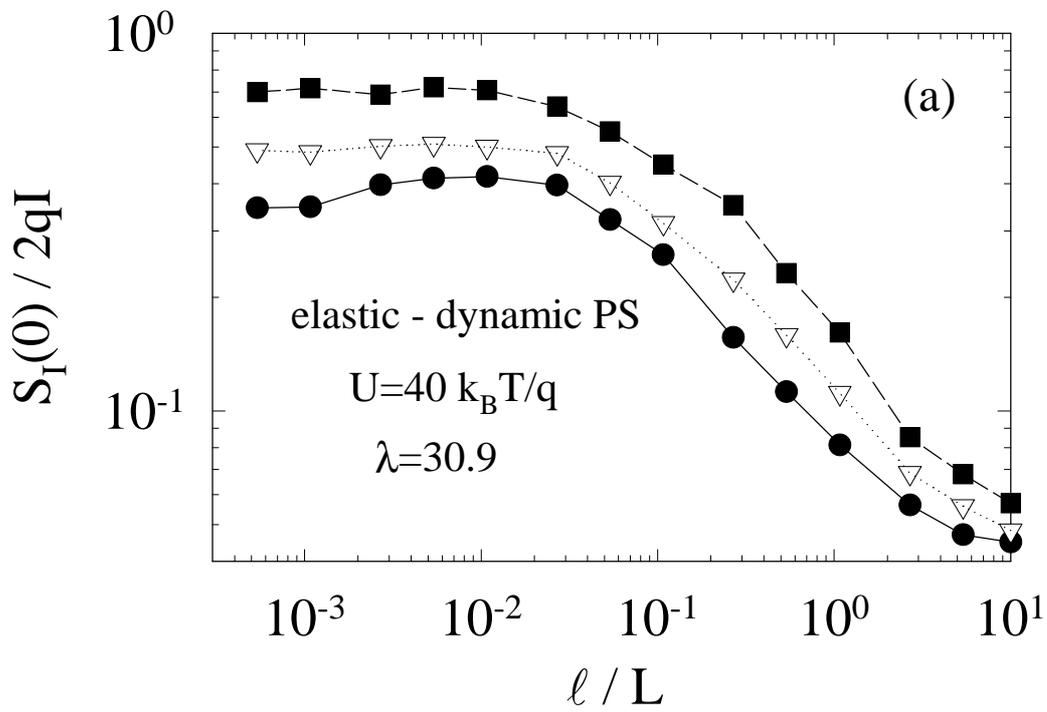
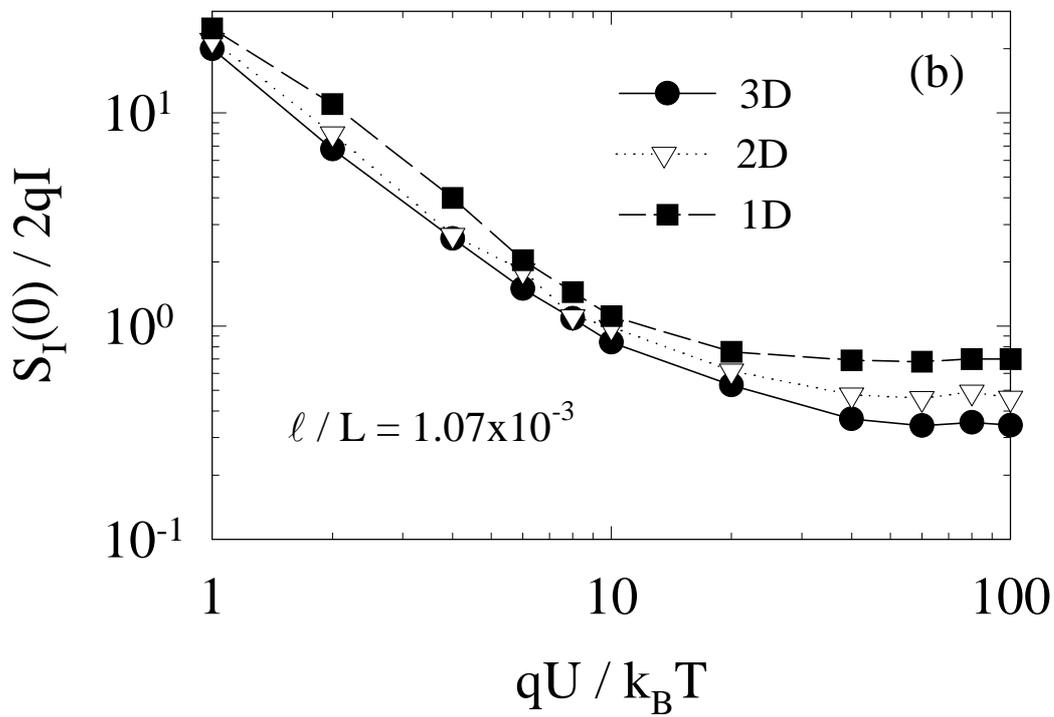
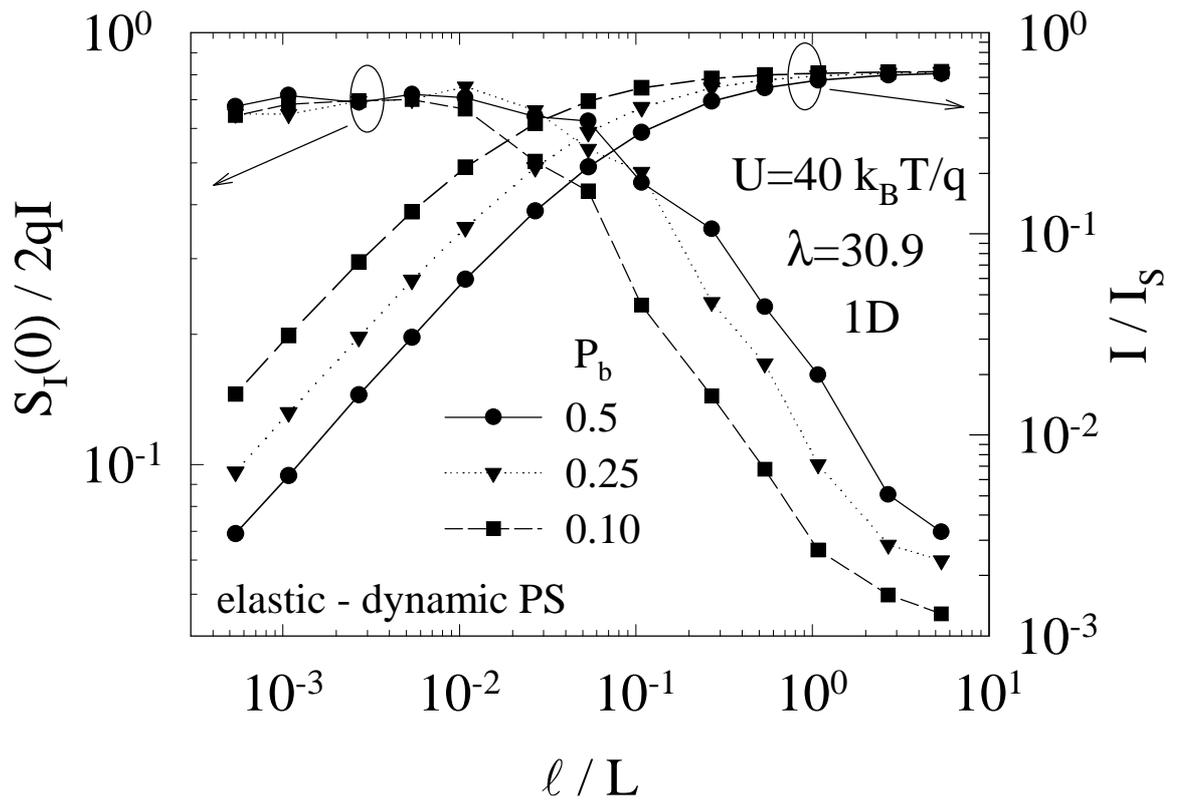
\end{document}